\documentclass[prl,aps,twocolumn,showpacs,preprintnumbers,amsmath,amssymb]{revtex4}

\usepackage{graphicx}
\usepackage{dcolumn}
\usepackage{bm}

\begin{document}


\title{H$_2$ dissociation over Au-nanowires and the fractional conductance quantum}

\author{Pavel Jel\'{\i}nek$^{1,2}$ \email{pavel.jelinek@uam.es},
Rub\'en P\'erez$^{1}$, Jos\'e Ortega$^{1}$ and Fernando Flores$^{1}$}

\affiliation{
$^{1}$ Departamento de F\'{\i}sica Te\'orica de la Materia Condensada,
Universidad Aut\'onoma de Madrid, E-28049 Spain}

\affiliation{
$^2$ Institute of Physics,
Academy of Sciences of the  Czech Republic,
Cukrovarnick\'a 10, 1862 53,Prague, Czech Republic}


\date{\today}

\begin{abstract}
The dissociation of H$_2$ molecules over stretched Au nanowires
and its effect on the
conductance are analyzed using a
combination of Density Functional (DFT) total energy calculations and
non-equilibrium Keldysh-Green function methods. Our
DFT simulations reproduce the characteristic
formation of Au monoatomic chains with a conductance close to
$G_0 = 2e^2/h$. These  stretched Au nanowires are shown to be
better catalysts for H$_2$ dissociation than Au surfaces. This is
confirmed by the nanowire conductance evidence: while insensitive to 
molecular hydrogen, atomic hydrogen
induces the appearance of fractional conductances ($G \sim 0.5
G_0$) as observed experimentally.
\end{abstract}

\maketitle

Gold surfaces are chemically inert and regarded as poor catalysts
at variance with other metal surfaces.
The low reactivity of molecular hydrogen on noble metal surfaces,
like Au and Cu, seems to be well understood
\cite{Norskov,Stromquist98,Brivio99,Lemaire02}. Density Functional
(DFT) calculations~\cite{Norskov} have shown that the dissociation
of $H_2$ on Au or Cu is an activated process: for large
molecule-surface distance, $d$, the interaction energy is
repulsive with a high barrier of 1.1 eV (around $d$ = 1.5 \AA),
which the molecule has to overcome to move along the reaction path
$ \rm H_{2} \to H+H$ and reach the final atomic chemisorption
state, with a total adsorption energy (2.07 eV per
atom~\cite{Lemaire02}) that is less than the $H_2$ binding energy
( 4.75 eV).
Compared to surfaces, small particles are known to be better
catalysts. Au is, in particular, considered as an exceptional
catalyst when prepared as nanoparticles on a variety of support
materials\cite{Haruta02}. Understanding this strong catalytic
activity is still the subject of an extensive research effort
with different possible explanations, including the particle shape
or perimeter, support effects and the metal oxidation state
\cite{Haruta02,Molina03,Goodman04,Yoon05}.

Nanowires are a~good example of systems whose size is so small
that one can expect their reactivity with molecules to be
increased considerably.
The formation of metallic nanocontacts has been analyzed in detail
thanks to the gentle control of the distance at the atomic scale
provided by both the Scanning Tunneling Microscope and the
Mecanically Controllable Break Junction~\cite{Rubio01,Scheer97}.
While, in many metallic contacts, the formation of nanowires
during the last stages of the stretching, just before the breaking
point, is characterized by an~atomic dimer geometry (Al is a
paradigmatic case~\cite{Jelinek03,Jelinek05}), in Au the final
geometry seems to be a chain of several atoms between the two
electrodes with a conductance close to the conductance quantum
unit $G_{o} = \frac{2e^2}{h}$ ~\cite{Yanson98}.
Although the formation and stability of monoatomic Au chains has
been addressed by several authors~\cite{Torres, Hakkinen,
daSilva01}, certain relevant aspects --in particular, the changes
in the structural and transport properties of the nanocontacts
induced by the presence of impurities~\cite{Bahn, Novaes, Legoas,
Barnett04}-- are not yet fully understood.

Recently , Csonka et al~\cite{Csonka}
have analyzed the interaction of  $H_2$ with a
breaking gold nanowire and have found new fractional peaks (in
units of  $G_{o}$) in the
conductance histogram. Moreover, conductance traces in a~stretched
nanowire demonstrate a~reversible transition between fractional
and integer conductances, in a~time scale of milliseconds  or
seconds, suggesting successive adsorption and desorption of
hydrogen on the chain.
These experiments do not show these effects for Cu and Ag, where
stable single-atom chains are not formed. This suggests the great
importance of the Au chains in the variation of gold conductance
in the presence of $H_2$.
Csonka et al~\cite{Csonka} have proposed a possible explanation
for the observed behaviour in terms of a dimerization effect which
has been theoretically predicted for idealized clean Au nanowires
~\cite{Hakkinen, Okamoto}. However this dimerization  has neither
been observed in conductance histogram measurements nor in more
complex theoretical simulations \cite{Sanchez-Portal, daSilva01,
Rubio01}.


\begin{figure}
\includegraphics[width=90mm,height=70mm]{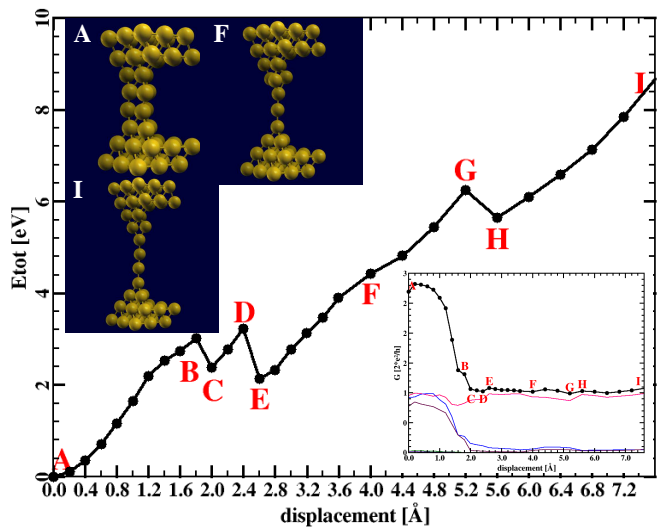}
\caption{\label{fig:etot} Total energy per unit cell and
ball-and-stick structure models (see also Figure 2)
for the Au nanowire as a function
of the stretching displacement. 
The inset shows the
total differential conductance 
and channel contribution along the stretching path.}

\vspace*{-0.5cm}

\end{figure}

In this Letter, we show that there is, in fact, a strong link
between the  enhanced reactivity of the stretched monoatomic gold
chains and the appearance of the fractional conductance peaks.
%
First, we have simulated the whole deformation process for an Au
nanocontact upon stretching, finding the formation and the final
breaking of a 4-atom Au chain.
The realistic nanocontact configurations, calculated in this way,
are then used to investigate whether the new fractional peaks in
the conductance 
are associated with adsorbed
molecular or atomic hydrogen. As our analysis suggests that only
atomic hydrogen can be responsible for these changes in the
nanowire conductance,  we have also investigated
how the chemical reaction $\rm H_{2} \to H+H$ is affected by the
presence of a freely suspended Au-wire.
%
This simplified model captures the key ingredients in the real
nanocontact structure and provides a natural playground to explore
the influence of the low dimensionality and the strain in the
different steps of the dissociation process.
These calculations show that a stretched Au-nanowire is
much more reactive than the Au surfaces, with a small activation
barrier, around 0.1 eV, for the H$_2$ dissociation and larger
chemisorption energies. Our results for the nanocontact
conductance, combined with the low value we have calculated for
the $H_2$-reaction activation barrier, strongly suggest
that the molecule dissociates on a Au-nanowire and that the
observed fractional conductance upon adsorption of molecular hydrogen is
basically due to the atomic hydrogen produced in the reaction.

%
%

Our calculations for stretched Au-nanowires have been performed
using a fast local-orbital DFT-LDA code (Fireball2004~\cite{Fir04}).
This code offers a very favorable
accuracy/efficiency balance if the basis of excited pseudoatomic
orbitals~\cite{basis} is chosen carefully.
%
The electrical conductance of the Au nanocontacts has been
calculated using a Keldysh-Green function approach based on the
first-principles tight-binding Hamiltonian obtained from the
Fireball code, at each point of the deformation path (see
references~\cite{Jelinek03,Jelinek05} for details).

First, we have analyzed the formation of a~Au-nanowire obtained by
stretching a~thick Au wire having four layers, with three atoms in
each layer, sandwiched between two (111)-oriented metal electrodes
as shown in figure~\ref{fig:etot} (configuration A represents the
initial relaxed configuration). We use a supercell approach, where
periodic boundary conditions  along different directions are
introduced: parallel to the surface we have considered a
3$\times$3-periodicity, while in the perpendicular direction we
join artificially the last layers of both electrodes(see
refs.~\cite{Jelinek03,Jelinek05}).
%
Figure~\ref{fig:etot} shows the total energy of the system as
a~function of the stretching displacement~\cite{relax_details}:
notice the energy jumps associated with the irreversible
deformations and structural rearrangements the wire has during
this process. This figure (see also fig. 2) shows several
snapshots for different geometries corresponding to the labels in
the energy curve. Our DFT simulations reproduce the characteristic
formation of Au monoatomic chains (with up to four atoms) found in
the experiments.
%
%
Similar results have been recently obtained using  parametrized
tight-binding molecular dynamics ~\cite{daSilva04}.


\begin{figure}
\hspace*{-5mm}
\includegraphics[width=30mm,height=25mm]{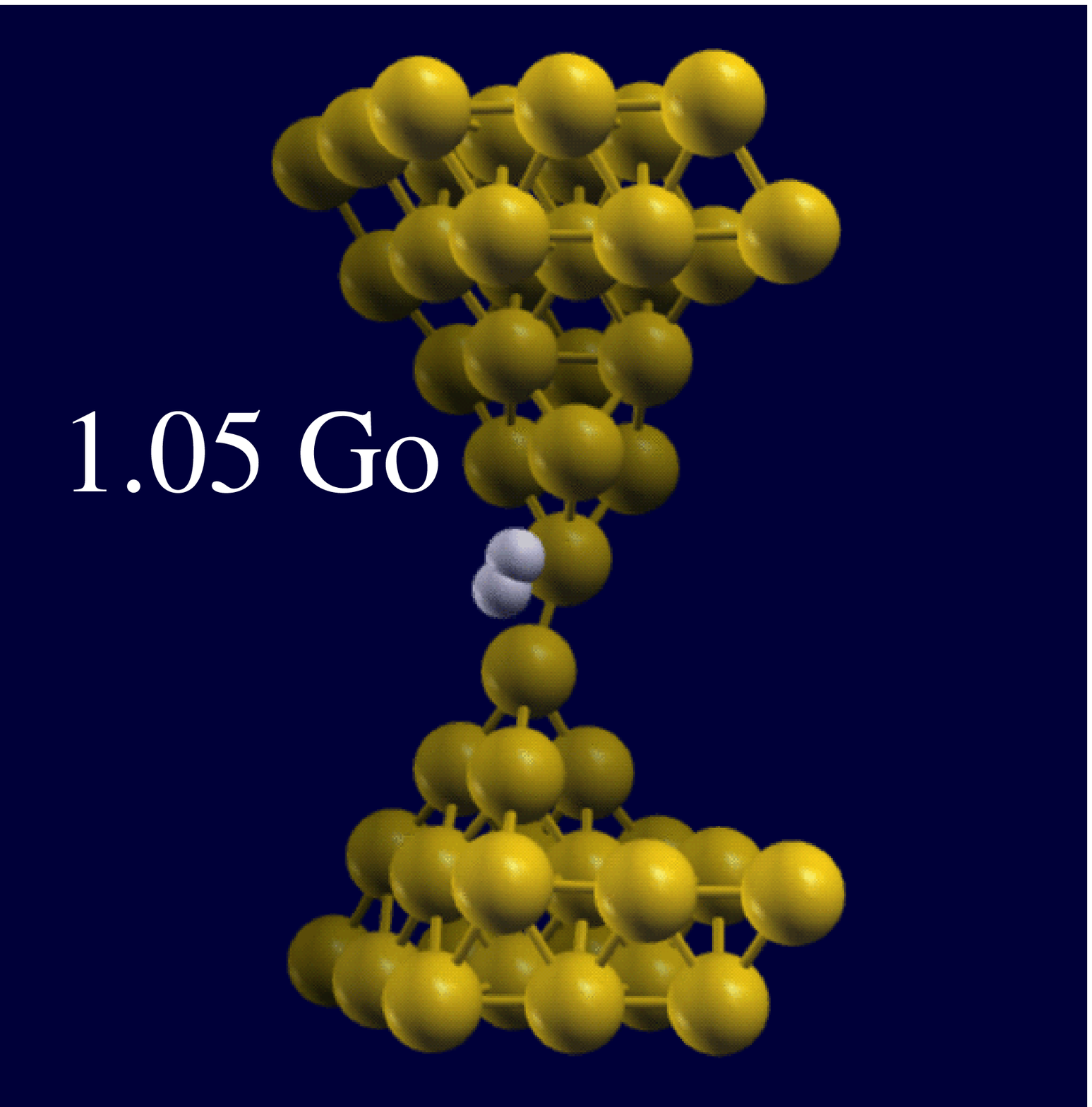}
\hspace*{-5mm}
\includegraphics[width=30mm,height=25mm]{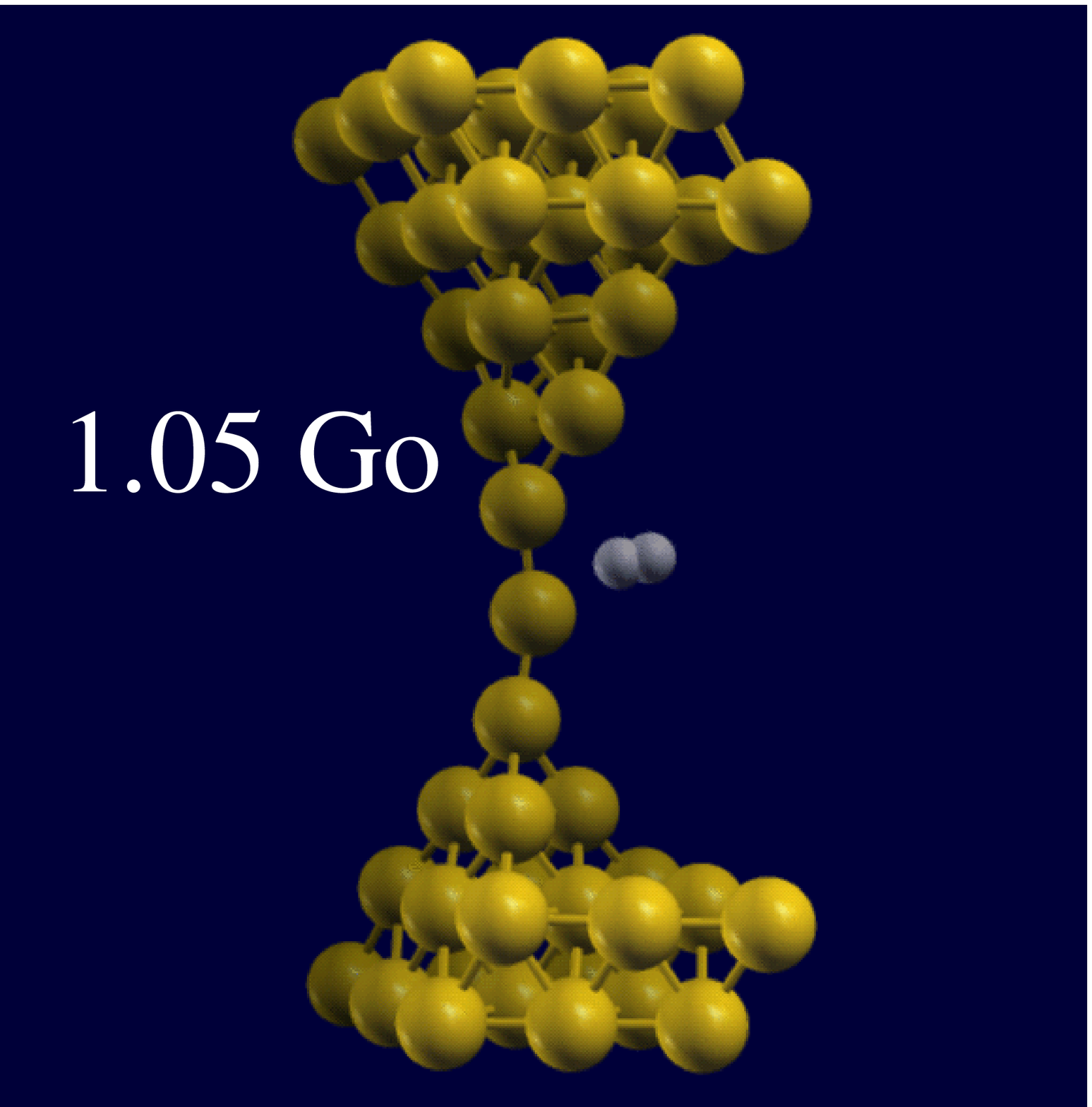}
\hspace*{-5mm}
\includegraphics[width=30mm,height=25mm]{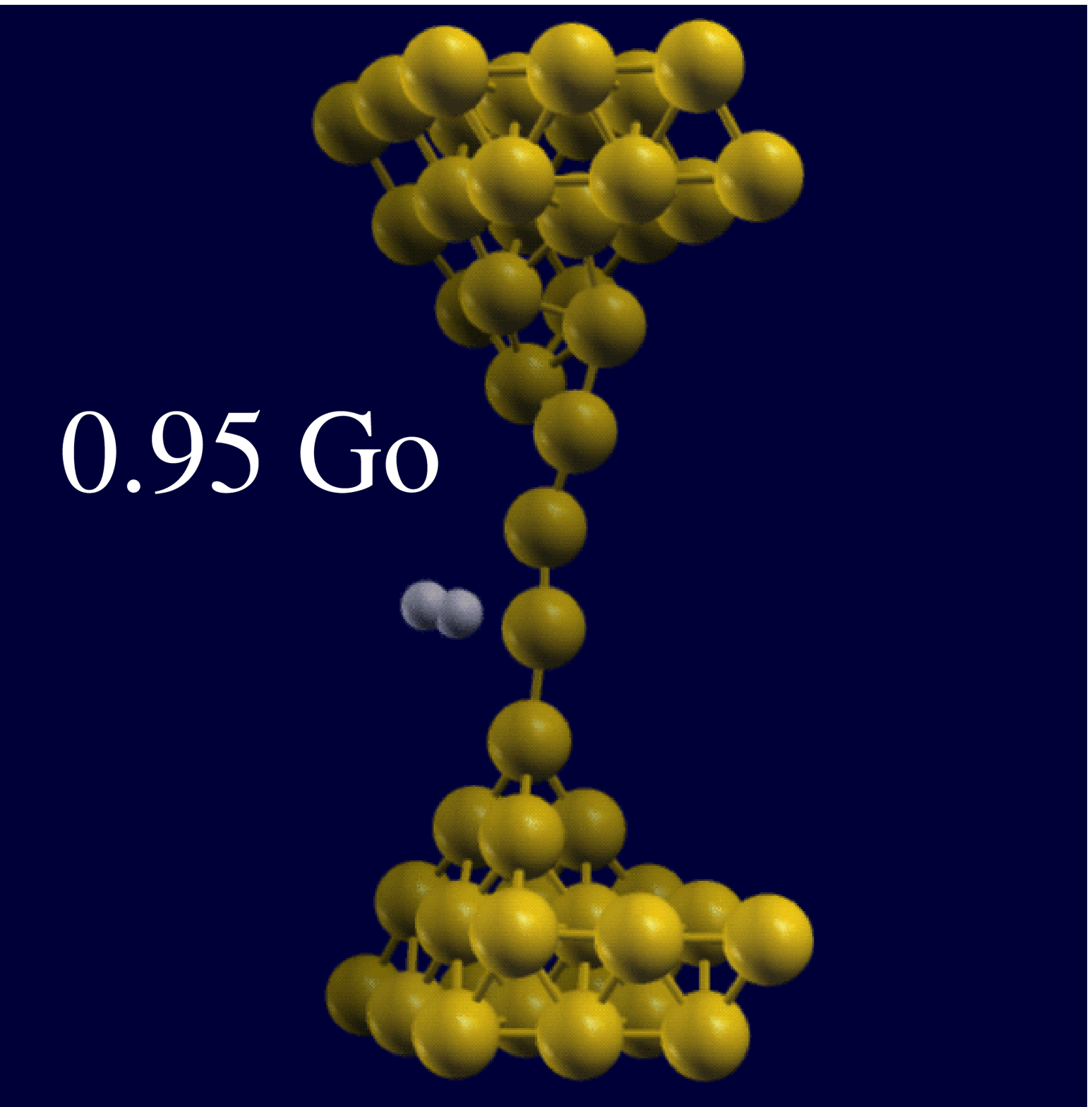}
\hspace*{-5mm}
\includegraphics[width=30mm,height=25mm]{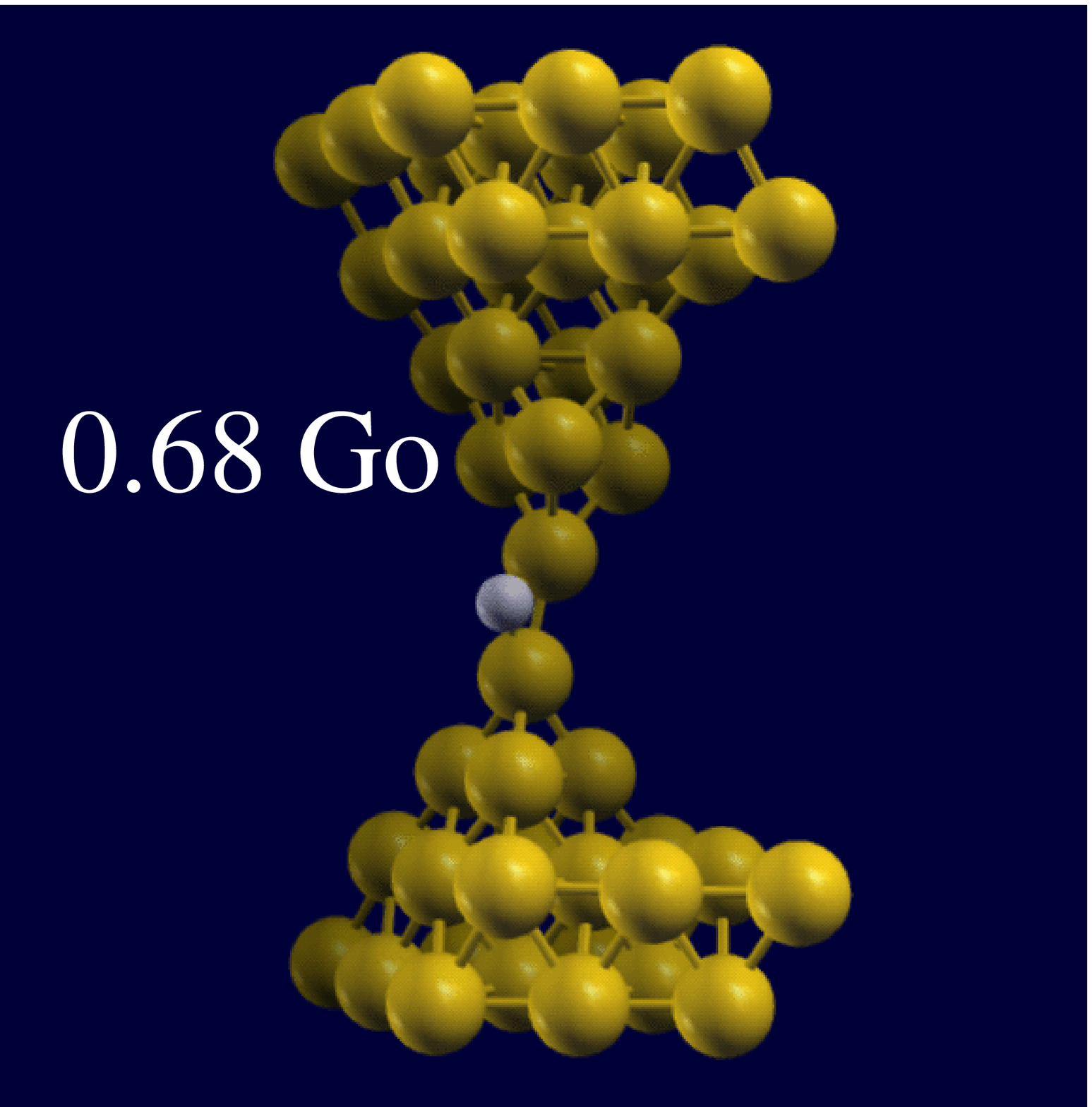}
\hspace*{-5mm}
\includegraphics[width=30mm,height=25mm]{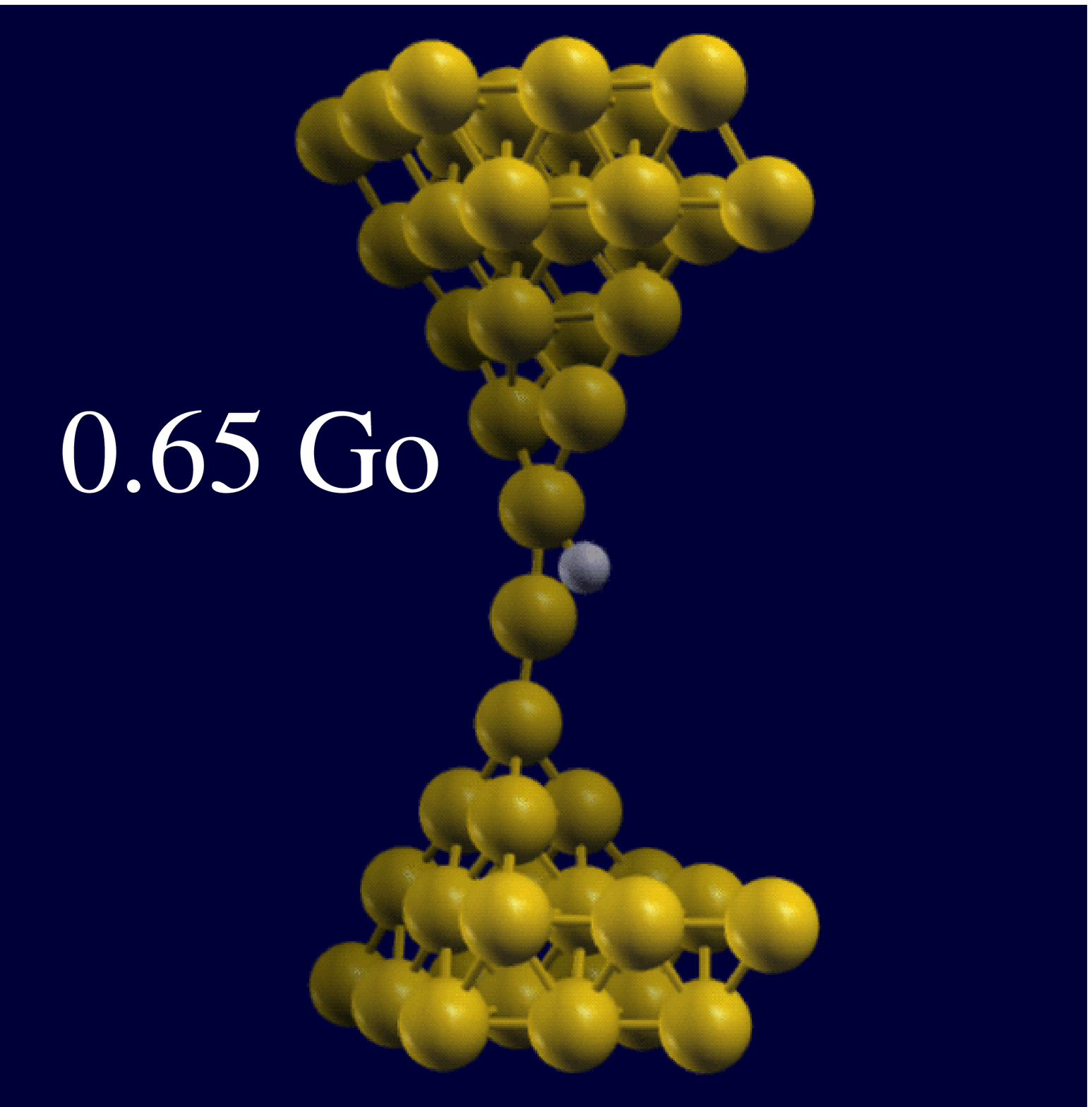}
\hspace*{-5mm}
\includegraphics[width=30mm,height=25mm]{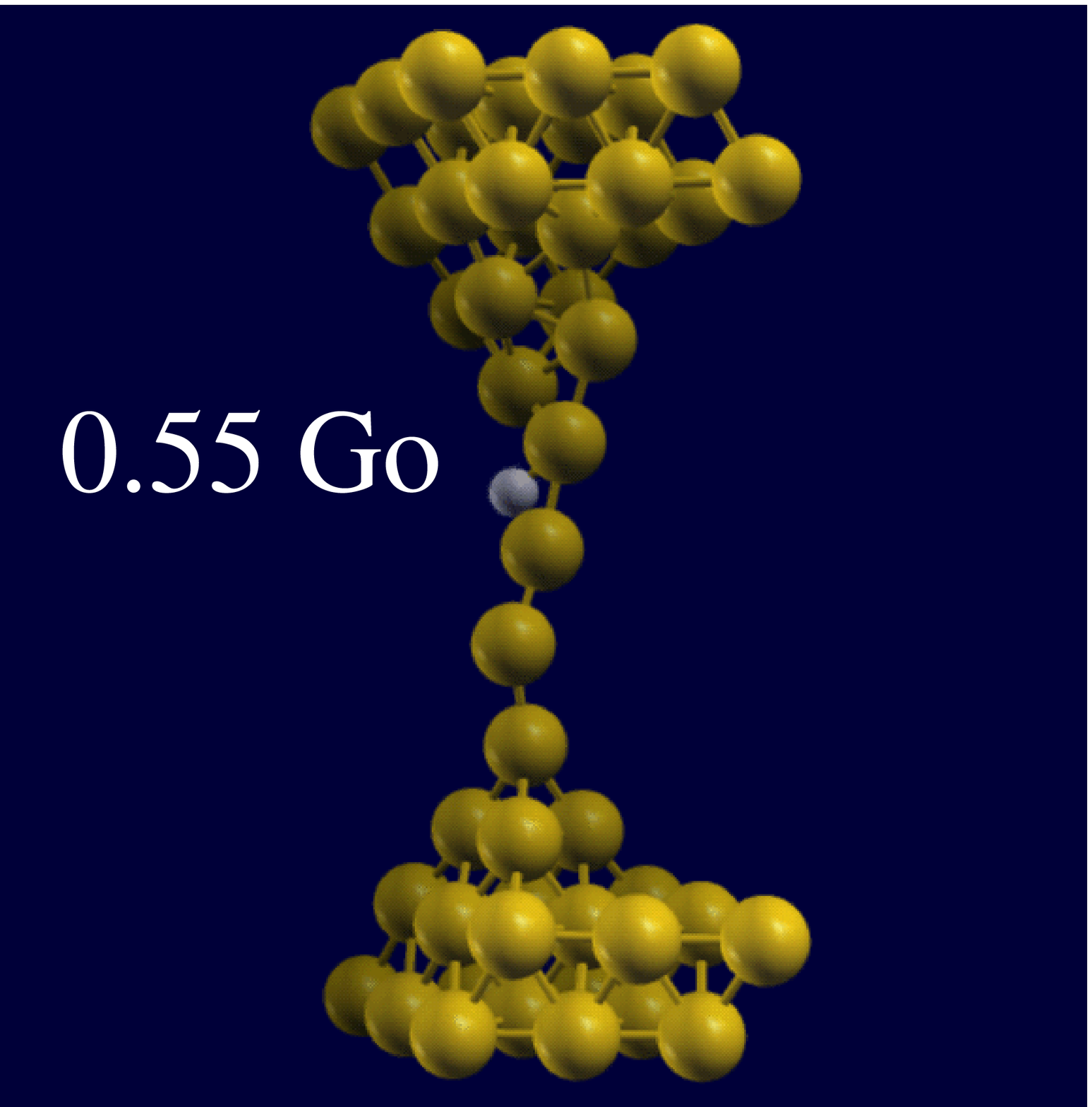}
\caption{\label{fig:wire} Ball-and-stick structure models and
total conductance for the adsorption of molecular (top) or atomic
hydrogen (bottom) on different nanowire geometries (corresponding
to the points E,F and G in Figure \protect\ref{fig:etot}).}

\vspace*{-0.5cm}

\end{figure}

The inset of figure~\ref{fig:etot} shows the conductance of the
system along the stretching process, as well as the different
channels contributing to it.
Notice that these results compare well with the experimental
evidence~\cite{Rubio01}. In particular, we reproduce (a) the long
conductance plateau associated with the formation of the
monoatomic chain (configurations C $\to$ I), where the conductance
is basically controlled by a~single channel associated mostly with
the Au s-electrons; and (b) the conductance oscillations during
the elongation process.
%
The very good agreement between  both our structural and
conductance results for the evolution of the Au nanocontact and
the experimental evidence provides strong support to the remaining 
simulations presented in this paper.

\begin{figure}
\includegraphics[width=70mm]{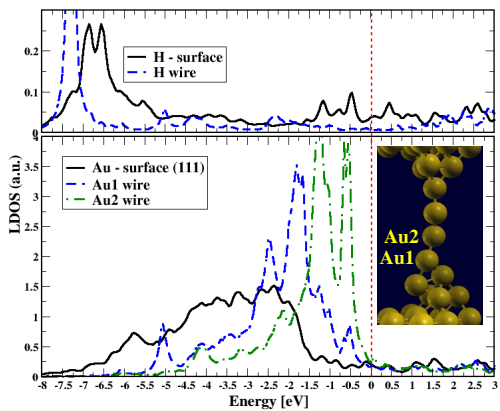}
\caption{\label{fig:pdos} (a) Density of states (DOS) for H
chemisorbed on the Au(111) surface (solid line) and on the Au
nanowire (dashed line). (b) Comparison of the DOS for topmost
atoms in the Au(111) surface and for Au atoms in the four atom
chain of the nanowire (see inset).}

\vspace*{-0.5cm}

\end{figure}

%
%
In a~second step, we have analyzed the nanowire conductance upon
the adsorption of molecular and atomic hydrogen. Starting with the
different geometries corresponding to the points E, F and G in
figure~\ref{fig:etot}, we have analyzed, via our local-orbital DFT
code,  how molecular and atomic hydrogen are adsorbed on those
geometries and, then, how the nanowire conductance is modified
according to the new optimized structures (see Figure
\ref{fig:wire}).
Molecular hydrogen is weakly adsorbed on all the different
nanowires, with energies of around 0.3~eV. Notice that, in all
these cases, molecular hydrogen does not penetrate the nanowire
too much (see figure~\ref{fig:wire}) 
and the conductance properties of the nanowire are not affected
practically by the adsorption of molecular hydrogen. In
particular, for the three cases shown in figure~\ref{fig:wire},
the nanowire conductance takes the values 1.05 $G_o$, 1.05 $G_o$
and 0.95 $G_o$, respectively. Atomic hydrogen introduces more
dramatic changes: in particular, for the cases shown in
figure~\ref{fig:wire}, the adsorption energies are 3.3 eV ,3.5 eV
and 3.9 eV for the chains with two, three or four atoms in the
nanowire, respectively. These values are much larger than the ones
found for H adsorbed on a~surface (around 2.1
eV)~\cite{Norskov,Lemaire02}.
Moreover, we find significant modifications in the conductance of
these three cases, with total values of 0.68 $G_{o}$, 0.65 $G_{o}$
and 0.55 $G_{o}$, respectively.
%
%
The eigenchannel analysis for all these cases shows that the
transport is dominated, as in the clean nanowire, by a channel
mostly associated with the Au s-electrons  but with a reduced
transmitivity. This reduction is related to the significant
displacement of the density of states (DOS) to lower energies
(particularly evident for the d bands) for the Au atoms bonded to
hydrogen that results in a reduction of the total DOS at E$_F$.
The nanowires with three or four atoms have been reanalyzed
assuming that two hydrogen atoms are simultaneously adsorbed on
the chain: the case of three atoms (case F) presents a~conductance
of 0.45 $G_{o}$, while for a~four-atom nanowire (case G) the
conductance is 0.2 $G_{o}$.

%
%
The enhanced reactivity of  Au-chains, with respect to
Au-surfaces, is due to the change in the Au DOS. Fig. 3b shows
this DOS for the atoms of the Au(111)-surface, and for the
Au-atoms of the four atom chain in the nanowire: Au-atoms with
lower coordination form directional bonds and present a narrower
DOS shifted towards the Fermi level. Then, for H chemisorbed on
the chain, its DOS (see fig. 3a) presents a tightly bound state at
-7.3 eV below the Fermi level, whereas the DOS for H on the
Au(111)-surface presents a broadened resonance with some
contribution from antibonding states just below the Fermi level
\cite{Norskov}.
Notice that the reactivity of Au chains is further increased by
the wire stretching due to a further shift of the Au-bands towards
the Fermi level.


\begin{figure}
\includegraphics[width=70mm]{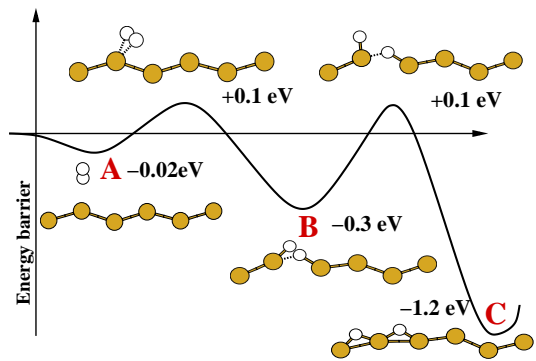}
\caption{\label{fig:barrier}Energetics of the H$_2$ dissociation
on a stretched Au nanowire along an arbitrary configurational
path.} 

\vspace*{-0.5cm}
\end{figure}

It is remarkable that, in the geometries shown in
figure~\ref{fig:wire}, neither the molecule nor the atom
penetrates the nanowire, breaking the bond between two Au atoms.
Recent work by Barnett et al~\cite{Barnett04} showed that, for an
essentially broken Au wire (with $G\sim 0.02 G_0$),  a barrierless
insertion of the H$_2$ molecule into the contact is possible. This
would correspond, in our case, to a wire state beyond the
configuration I in figure~\ref{fig:etot}, where the conductance is
still very close to $G_o$. They then used this configuration as a
starting point for a detailed study of the structure, orientation
and stability of the molecule upon compression of the wire.
Our approach is different, since we are interested in the
wire-H$_2$ interaction during the chain formation process. We have
calculated the energy barriers the molecule experiences when
moving from the geometries shown in figure~\ref{fig:wire} to
inserted sites. Our calculations yield values larger than 0.5 eV.

%
%
Both the conductance and the total energy results discussed so far
support the dissociation of molecular hydrogen in the Au
monoatomic chains. As the energetics were discussed in terms of
the local orbital code, using the LDA approximation for the
exchange-correlation, we have re-analyzed the dissociation
mechanism on freely suspended Au-wires, with H (or $H_{2}$)
adsorbed on them, using CASTEP ~\cite{castep,technical}, with a
gradient corrected approximation~\cite{Perdew91} (GGA) for the
exchange-correlation functional.
This simplified model for the nanowire is dictated by the
computational resources needed for this full calculation,
but it offers the possibility to discuss the
relative contribution of the low dimensionality and the strain in
the enhanced reactivity of the nanowire.

We have considered a freely suspended Au wire with six independent
atoms per chain and periodic boundary conditions chosen to produce
a stretch deformation in order to simulate the breaking process.
After relaxing the free wire for each strain condition, we have
studied how molecular hydrogen interacts with it: in particular,
we have explored the possibility of having the following reaction,
$\rm H_{2} \to H+H$ on the suspended wire.
Figure~\ref{fig:barrier} shows our main results for a wire with a
11\% strain in an~arbitrary configurational path: the case A in
the figure represents the geometry of $\rm H_{2}$ interacting with
the freely suspended wire; this is a physisorbed state with
an~adsorption energy of around 0.02 eV. Case B corresponds to
an~intermediate state in which a~H-atom is chemisorbed  between
two Au-atoms , and the other H is still bonded to the Au-atom on
which the molecule was initially physisorbed: this case has
an~adsorption energy of 0.3 eV; we have found, however, a~barrier
of 0.1 eV between states A and B . Case C corresponds to the final
reaction state, in which the two H are adsorbed between two
Au-atoms; the chemisorption energy of this final state is 1.4 eV,
but again we find an energy barrier of 0.4 eV between states B and
C.
The reaction path drawn in figure~\ref{fig:barrier} shows that
molecular hydrogen sees a~total barrier of 0.1 eV for the reaction
$\rm H_{2} \to H+H $ on a~freely suspended Au-wire. In
a~Au-surface, the energy barrier for that reaction is around 1.0
eV, and the adsorption energy,  with respect to $H_{2}$ , is
negative, around 0.5 eV. These numbers show the extreme importance
that a stretched Au-nanowire has in the reactivity of molecular
hydrogen: its reaction energy barrier and its chemisortion energy
are lowered at least by 1 eV by the nanowire.
Notice that the reduced dimensionality of the nanowire (compared
to the surface) is not enough to induced this high reactivity, as
shown by similar calculations for a non-strained chain, where we
find an initial bound state for the molecule with energy -0.10 eV
and quite large barriers ($\sim 0.5$ eV) for the dissociation.



In conclusion, stretched Au nanowires are much more reactive with
molecular hydrogen than Au surfaces. Our DFT-GGA calculations show
that, in a stretched Au nanowire, the activation barrier for H$_2$
dissociation is very small, around 0.10 eV.
Notice that this is a reliable upper bound, as the use of the the
GGA approximation (that only partially corrects the gross
overestimation of barriers in LDA), the freely suspended wire
geometry considered, and the neglect of possible quantum tunneling
effects in H tend to overestimate the calculated barrier.
This enhanced reactivity can also be expected in the case of Pt,
where the formation of chains of several atoms has been also
observed \cite{Smit01}.
Complementary evidence is given by Csonka et al~\cite{Csonka} data
for the conductance of a stretched Au nanowire with adsorbed
molecular hydrogen. Our calculations indicate that only atomic
hydrogen can be responsible of the changes observed by those
researchers in the nanowire conductance. It is the combination of
these two results, our calculated activation barrier and the
changes in the nanowire conductance, that strongly suggest that
molecular hydrogen dissociates when adsorbed on stretched Au
nanowires.
%



\begin{acknowledgments}
P.J. gratefully acknowledges financial support by the Ministerio
de Educacion y Ciencia of Spain and the Ministry of Education,
Youth and Sports of Czech Republic. This work has been supported
by the DGI-MCyT (Spain) under contracts MAT2002-01534 and
MAT2004-01271. Part of these calculations have been performed in
the Centro de Computaci\'on Cient\'{\i}fica de la UAM. We thank
Dr. J.C. Conesa and Prof. Sidney Davidson for helpful comments.
\end{acknowledgments}


\vspace*{-0.5cm}

\begin{thebibliography}{33}

\vspace*{-0.5cm}

\bibitem{Norskov}
B. Hammer, J.K. Norskov, Nature {\bf 376}, 238 (1995).

\bibitem{Stromquist98}
J. Stromquist et al, Surf. Sci.  {\bf 397}, 382 (1998).

\bibitem{Brivio99}
G.P. Brivio and M.I. Trioni, Rev. Mod. Phys. {\bf 71}, 231 (1999).

\bibitem{Lemaire02}
D. Lemaire, J.G. Anattrucci and B. Jackson, Phys. Rev. Lett. {\bf
89}, 268302 (2002).

\bibitem{Haruta02}
M. Haruta, Cat. Tech. {\bf 6}, 102 (2002).

\bibitem{Molina03}
L.M. Molina and B. Hammer, Phys. Rev. Lett. {\bf 90}, 206102
(2003).

\bibitem{Goodman04}
M.S. Chen and D.W. Goodman, Science {\bf 306}, 252 (2004).

\bibitem{Yoon05}
B. Yoon et al., Science {\bf 307}, 403 (2005).


\bibitem{Rubio01}
 G. Rubio-Bollinger, S. R. Bahn, N. Agr\"{a}it, K. W. Jacobsen, and S. Vieira,
Phys. Rev. Lett. {\bf 87}, 026101 (2001).

\bibitem{Scheer97}
E. Scheer et al, Phys. Rev. Lett. {\bf 78}, 3535 (1997).

\bibitem{Jelinek03}
P. Jelinek et al, 
Phys. Rev. B {\bf 68}, 085403 (2003).

\bibitem{Jelinek05}
P. Jelinek et al, 
Nanotechnology {\bf 16}, 1023 (2005).


\bibitem{Yanson98}
A.I. Yanson et al, Nature {\bf 395}, 783 (1998).

\bibitem{Hakkinen}
H. H\"{a}kkinen, R.N. Barnett and U. Landman, J. Phys. Chem B {\bf
103}, 8814 (1999).

\bibitem{Torres}
J. A. Torres et al., Surf. Sci. {\bf 426}, L441 (1999).
\bibitem{daSilva01}
E. Z. da Silva, A. J. R. da Silva, and A. Fazzio, Phys. Rev. Lett.
{\bf 87}, 256102 (2001).

\bibitem{Barnett04}
R. Barnett et al, NanoLetters {\bf 4}, 1845 (2004).

 \bibitem{Bahn}
S.R. Bahn, N. Lopez, J.K. Norskov and K.W. Jacobsen, Phys. Rev. B
{\bf 66}, 081405 (2002).
 \bibitem {Novaes}
F. D. Novaes, A. J. R. da Silva, E. Z. da Silva,and A. Fazzio,
Phys. Rev. Lett. {\bf 90}, 036101 (2003).
\bibitem{Legoas}
S. B. Legoas, D. S. Galvão, V. Rodrigues, and D. Ugarte
 Phys. Rev. Lett. {\bf 88}, 076105 (2002).

\bibitem{Csonka}
Sz. Csonka et al, 
Phys. Rev. Lett. {\bf 90}, 116803 (2003).

\bibitem{Okamoto}
M. Okamoto, K. Takayanagi, Phys. Rev. B {\bf 60}, 7808 (1999)

\bibitem{Sanchez-Portal}
D. S\'anchez-Portal  et al, Phys. Rev. Lett. {\bf 83}, 3884
(1999).

\bibitem{Fir04}
P. Jelinek et al,
Phys. Rev. B {\bf 71},  235101 (2005);
J.P. Lewis et al,
Phys. Rev. B {\bf 64},  195103 (2001).

\bibitem{basis}
For Au, the valence electrons were described by $s,p,d$ slightly
excited pseudoatomic orbitals~\cite{Sankey}. For the cutoff radii
$\rm R_c(s) = 4.5 a.u., \rm R_c(p) = 5.2 a.u., \rm R_c(d) =  a.u.$
we obtained for f.c.c. Au a~lattice parameter of a = 4.14 \AA\ and
a bulk modulus of B = 210 GPa (experiment: a = 4.07 \AA, B = 173
GPa). For H, we used double numerical $s$-orbital basis set with
cutoff radii $\rm R_c(s)= 3.8 a.u.$.


\bibitem{relax_details}
The nanowire is stretched by increasing the distance between the
upper and lower fixed layers by 0.2 A at each step and letting the
rest of the atoms to relax towards its ground state configuration.

\bibitem{daSilva04}
E.Z. da Silva et al , Phys. Rev. B {\bf 69}, 115411 (2004).

\bibitem{castep}
CASTEP 4.2 Academic version, licensed under the UKCP-MSI Agreement, 1999.
M.C. Payne et al., Rev. Mod. Phys. {\bf 64}, 1045 (1992)

\bibitem{technical}
Au and H atoms are described with ultrasoft
pseudopotentials~\protect\cite{vanderbilt} and wavefunctions
expanded with a PW cutoff of 340 eV.

\bibitem{Perdew91}
J.P. Perdew et al, 
Phys. Rev. B {\bf 46}, 6671 (1992).

\bibitem{vanderbilt}
D. Vanderbilt, Phys. Rev. B {\bf 41}, 7892 (1990).




\bibitem{Smit01}
R.H.M. Smit et al, Phys. Rev. Lett. {\bf 87}, 266102 (2001).


\end{thebibliography}
\end{document}